\documentclass{article}

\usepackage{spconf,amsmath,graphicx}
\usepackage{amsmath,amssymb,amsfonts}
\usepackage{cite}
\usepackage{algorithmic}
\usepackage{textcomp}

\usepackage[colorlinks=true, allcolors=blue]{hyperref}

\usepackage{mwe} 
\usepackage{multirow}
\usepackage{makecell}
\usepackage{adjustbox}
\usepackage{xcolor}
\usepackage{float}
\usepackage{subcaption}

\def\code#1{\texttt{#1}}

\begin{document}

\title{Automated Measurement of Pericoronary Adipose Tissue \\ 
Attenuation and Volume in CT Angiography}


\name{Andrew M. Nguyen, Tejas Sudharshan Mathai, Liangchen Liu, Jianfei Liu, Ronald M. Summers}
\address{Imaging Biomarkers and Computer-Aided Diagnosis Laboratory, \\ Radiology and Imaging Sciences, National Institutes of Health Clinical Center, Bethesda, MD, USA}

\maketitle

\begin{abstract}

\noindent 
Pericoronary adipose tissue (PCAT) is the deposition of fat in the vicinity of the coronary arteries. It is an indicator of coronary inflammation and associated with coronary artery disease. Non-invasive coronary CT angiography (CCTA) is presently used to obtain measures of the thickness, volume, and attenuation of fat deposition. However, prior works solely focus on measuring PCAT using semi-automated approaches at the right coronary artery (RCA) over the left coronary artery (LCA). In this pilot work, we developed a fully automated approach for the measurement of PCAT mean attenuation and volume in the region around both coronary arteries. First, we used a large subset of patients from the public ImageCAS dataset (n = 735) to train a 3D full resolution nnUNet to segment LCA and RCA. Then, we automatically measured PCAT in the surrounding arterial regions. We evaluated our method on a held-out test set of patients (n = 183) from the same dataset. A mean Dice score of 83\% and PCAT attenuation of -73.81 ± 12.69 HU was calculated for the RCA, while a mean Dice score of 81\% and PCAT attenuation of -77.51 ± 7.94 HU was computed for the LCA. To the best of our knowledge, we are the first to develop a fully automated method to measure PCAT attenuation and volume at both the RCA and LCA. Our work underscores how automated PCAT measurement holds promise as a biomarker for identification of inflammation and cardiac disease.


\end{abstract}

\section{Introduction}

Coronary artery disease is the leading cause of mortality worldwide, originating from the formation of atherosclerotic plaque within the coronary arteries \cite{Malakar2019_reviewOfCAD}. Studies have shown that the rupture of non-obstructive and highly inflamed plaques have led to $\sim$60\% of all myocardial infarctions (heart attacks) \cite{Ma2023_PCAT,Antoniades2019}. Therefore, finding biomarkers that can indicate inflammation and can be used as a predictor for the risk of cardiac disease are of clinical interest. In recent years, the fat deposition around the coronary arteries called pericoronary adipose tissue (PCAT) has been identified as a potential biomarker for early diagnosis and intervention \cite{Ma2023_PCAT,Antoniades2019,Oikonomou2018_coronaryInflamm,Goeller2018_PCAT,Chatterjee2022}. 


Presently, only semi-automated methods targeted towards PCAT measurement exist \cite{ALMEIDA2020490,Chatterjee2022,Xu2020_PCATradiodensity}. Furthermore, a disproportionate number of past research has focused on the RCA at the expense of the LCA for evaluation of PCAT. Measurement of PCAT surrounding the left main (LM), left anterior descending (LAD), and left circumflex (LCX) arteries may offer a broader perspective with respect to the distinct hormonal signaling pathways between fat and plaque \cite{Ma2023_PCAT}. Given the morphological and functional variability in the coronary arteries and their branches \cite{Goeller2018_PCAT,Oikonomou2018_coronaryInflamm,Ma2023_PCAT}, the development of a fully automated method for PCAT attenuation and volume measurements across all these arteries remains elusive. 



\begin{figure*}[!h]
\centering
\includegraphics[page=1, width=0.95\linewidth]{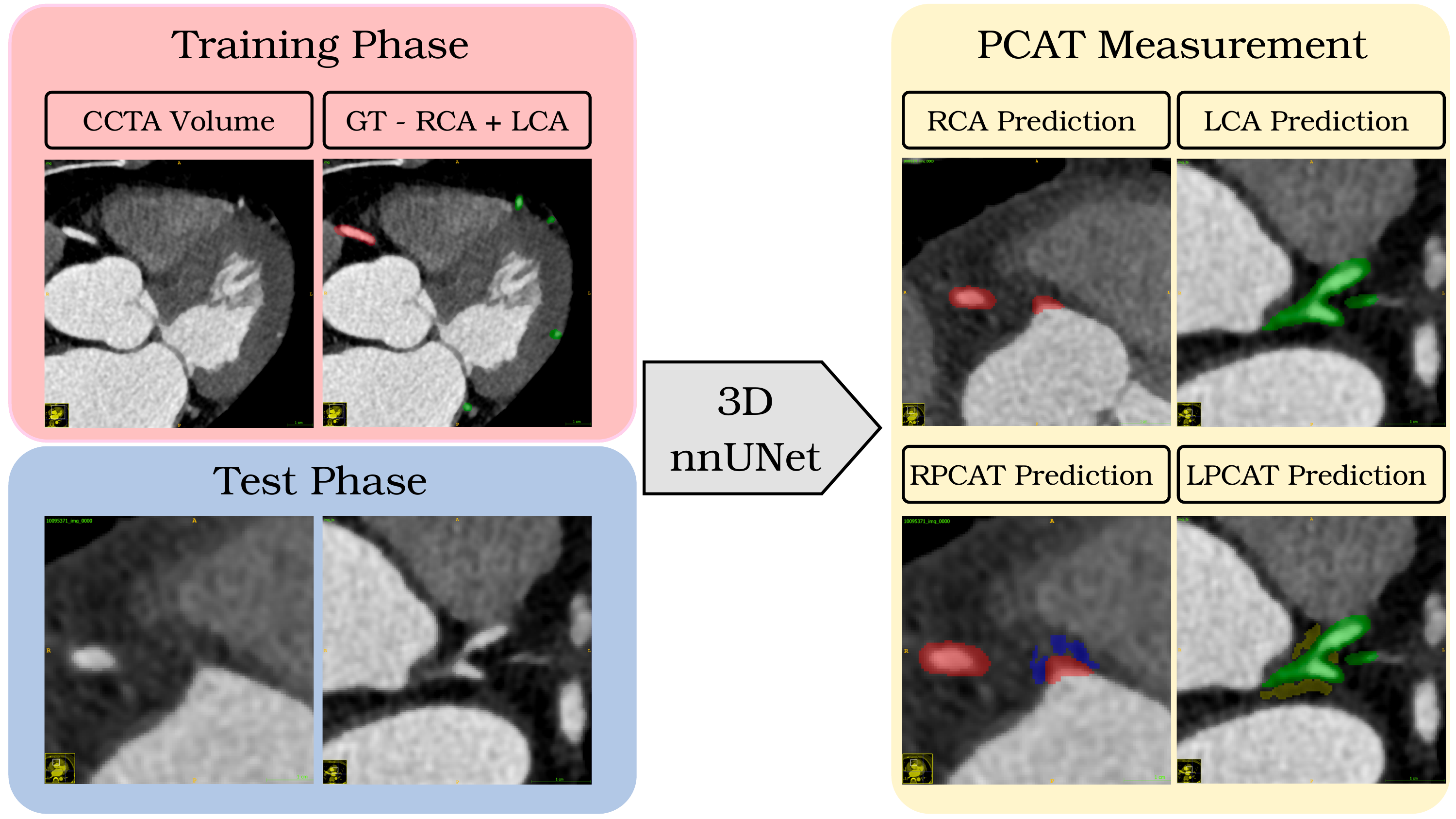}
\caption{Summary of automated measurement of pericoronary adipose tissue (PCAT) in coronary CT angiography (CCTA) volumes. In the training phase, a 3D nnUNet framework was trained on CCTA volumes and ground truth (GT) right coronary artery (RCA, red) and left coronary artery (LCA, green) masks. At inference time, the model was run on volumes from the test dataset. Finally, PCAT was automatically measured at the RCA (RPCAT, blue) and PCAT at the LCA (LPCAT, yellow).}
\label{fig_money}
\end{figure*}

In this pilot work, we developed a fully automated approach for the measurement of PCAT mean attenuation and volume in the region around both left and right coronary arteries. We used a large subset of patients from the public ImageCAS dataset (n = 735) to train a 3D full resolution nnUNet model segmentation of the LCA and RCA. Next, we automatically measured the PCAT in the vicinity of the RCA and LCA. We evaluated our pipeline on a held-out test set of patients (n = 183) from the same dataset.


\section{Methods}

\subsection{Data}

The ImageCAS dataset \cite{Zeng2023_ImageCAS} consists of 1000 3D CCTA volumes acquired from 1000 patients. There were 586 males and 414 females with average ages of 60 and 57 years, respectively. These patients had a history of ischemic stroke, ischemic heart attack and/or peripheral artery disease. A single Siemens CT scanner (128-slice, dual-source) was used for imaging, and the coronary arteries in this dataset were labeled by three radiologists. The labels consisted of the right coronary artery (RCA), left main coronary artery (LM), left anterior descending artery (LAD), left circumflex artery (LCX), and  other branches. These arteries branched out of the aorta, but the label mask for the aorta was not provided in this dataset. We segmented the aorta with the help of a publicly available tool called TotalSegmentator \cite{Wasserthal2023_TotalSegmentator}. 

\begin{figure}[!t]
    \centering
    \includegraphics[page=1, width=\columnwidth]{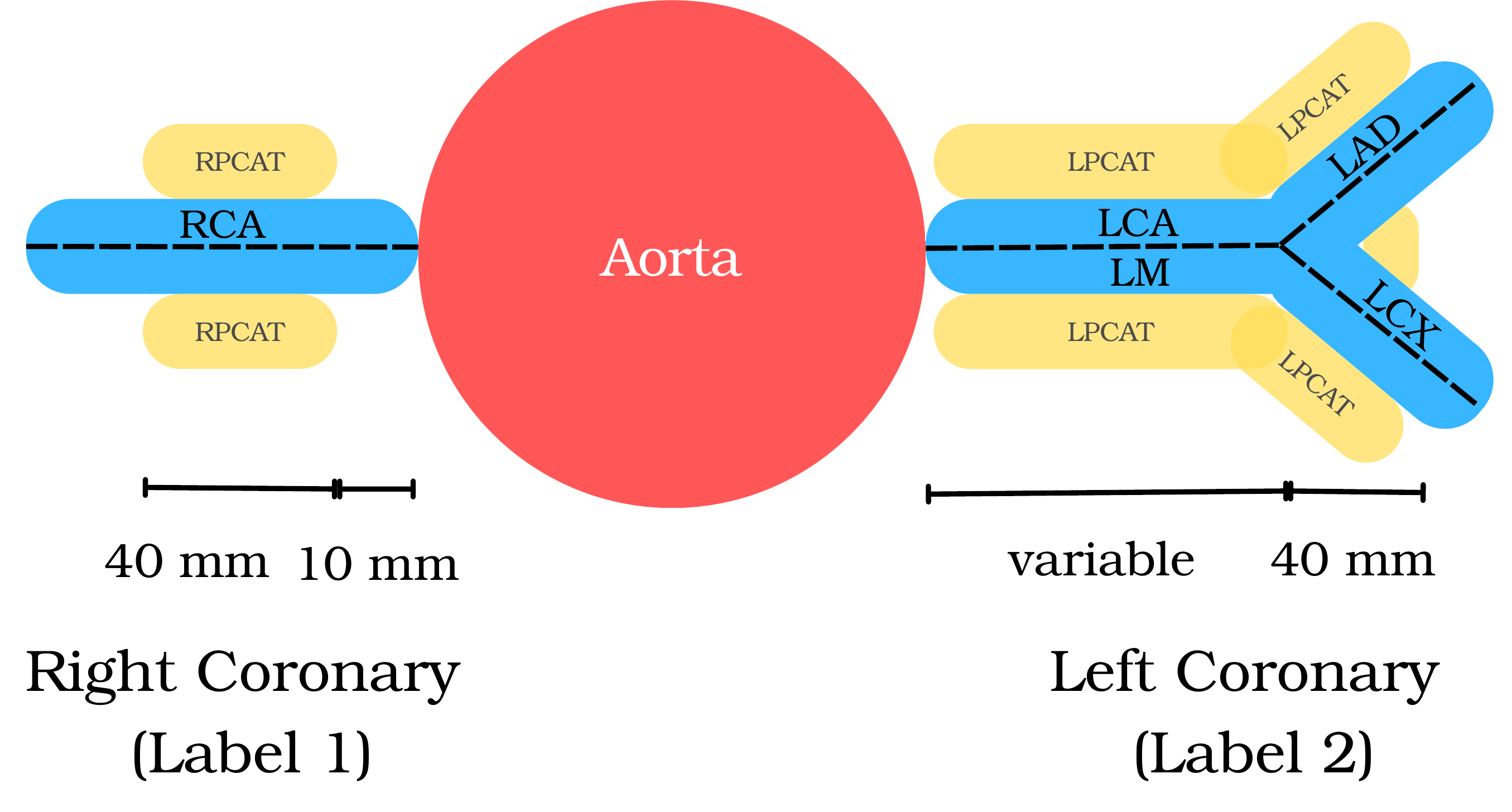}
    \caption{Illustration of the automated pericoronary adipose tissue (PCAT) measurement at the right coronary artery (RCA) and left coronary artery (LCA), which includes the left main (LM), left anterior descending (LAD), and left circumflex (CX) arteries. Dashed centerlines (black) were extracted from each artery (cyan). They were used to determine the segment length for PCAT measurement, and the bifurcation point of the LM into the LCX + LAD. PCAT at the RCA (RPCAT) was measured 10 mm distal to the ostium and for a 40 mm segment length. PCAT at the LCA (LPCAT) was measured along the full length of the LM, but only for a 40 mm segment from the bifurcation point.}
    \label{fig_arterySeparation_with_measurePCAT}
\end{figure}

Next, we ran connected component analysis to separate the arteries on the right side of the heart from the left side. As shown in Figs. \ref{fig_money} and \ref{fig_arterySeparation_with_measurePCAT}, we obtained two distinct labels; one for the RCA (label 1), and another for the LCA (label 2). The LCA consisted of the combined masks of the LM, LAD, and LCX. This resulted in 918 patient volumes with an effective split of the coronary arteries into LCA and RCA, and the exclusion of 82 volumes where the connected component analysis failed. This division was manually verified, and the dataset was split into a training set (n = 735 studies) and a held-out test set (n = 183 studies).

\subsection{Model}

We trained a 3D full resolution nnUNet model \cite{Isensee_2020} to segment the left and right coronary arteries. The nnUNet model is a self-configuring segmentation framework that can adapt to various datasets and modalities, including CT. It is currently the \textit{de-facto} standard for segmentation tasks. It automatically determines the optimal hyper-parameters for training a segmentation model and learns to segment target structures of interest. During training, our 3D full-resolution nnUNet took input CCTA volumes and the corresponding ground-truth
masks for the LCA and RCA (n = 735). The model learned to segment the target structures of interest in the CT volume and iteratively refined it via a loss function. This loss function computed a segmentation error that measured the overlap between the prediction and ground-truth. 

The framework executed a 5-fold cross-validation scheme, and used different subsets of training and validation data for training the model in each fold. At inference time, the 3D nnUNet predicted the segmentation masks for the LCA and RCA in the held-out test CT volumes. The best model with the lowest loss from each of the 5 folds was used for inference on the test CT volume, and predictions from these five folds were ensembled together (via majority voting). 

\subsection{Measurement of PCAT}

Once the LCA and RCA were segmented, the PCAT regions around them were measured. The perivascular fat region is defined as the adipose tissue within a 5 mm radial distance from the outer vessel wall that is within the [-190, -30] Hounsfield units (HU) range. To identify this region, the centerlines for the LCA and RCA were first extracted from the corresponding segmentations. We used the \code{skeletonize} function in Python \cite{pedregosa2011scikit} for this task. 


\smallskip
\noindent
\textbf{Right PCAT (RPCAT).} Consistent with prior work \cite{Xu2020_PCATradiodensity,Chatterjee2022}, we commenced the PCAT measurement at a distance of 10 mm distal to the ostium to avoid proximity to the aortic wall. As seen in Fig. \ref{fig_arterySeparation_with_measurePCAT}, at each point on the centerline of the RCA, we estimated the closest distance to the vessel wall. From this distance, the approximate diameter of the vessel was ascertained. Then, a sphere of radius equal to the vessel diameter was placed on each point along the centerline, and this resulted in a 3D cylinder that encompassed the region of the PCAT and the segmented vessel. All the voxels that fell within the aforementioned HU range were averaged together to provide the mean attenuation of the RPCAT. The RPCAT volume was computed by multiplying the number of voxels by the volumetric spacing.  



\smallskip
\noindent
\textbf{Left PCAT (LPCAT).} To locate the left PCAT area, we first identified the primary bifurcation point closest to the aorta that indicated the end of the LM artery and start of the LAD and LCX. The centerline voxel that had 3 or more neighboring centerline voxels was considered the bifurcation point. As shown in Fig. \ref{fig_arterySeparation_with_measurePCAT}, without the location of the primary bifurcation, the length of the segment to measure the PCAT along the LAD and LCX would be unknown. From this bifurcation point, a 40 mm segment was considered, such that the LCX and LAD were included. With these segments ascertained, we estimated the vessel diameter at each point along the centerlines for the LM, LAD, and LCX. Similar to the previous section, a sphere with radius equal to the vessel diameter was placed at each ceneterline voxel to obtain the PCAT region. The voxels in the specified HU range within this region were used to calculate the LPCAT mean attenutation and volume. 



\section{Experiments \& Results}

\subsection{Implementation}

The 3D nnUNet model was trained using 5-fold cross-validation with the different initialization of trainable parameters for a total of 1000 epochs. The loss function used by the model was an equally weighted combination of binary cross-entropy and soft Dice losses. It was optimized using the Stochastic Gradient Descent (SGD) optimizer with an initial learning rate of $10^{-3}$ and a batch size of 1. Each CT volume in the test split was passed to the model from each fold, and predictions from five folds were ensembled together. All experiments were done on a workstation running Ubuntu 22.04 LTS with a NVIDIA Tesla V100 GPU.

\subsection{Results}

\begin{figure}[!h]
\centering
\begin{subfigure}{0.33\columnwidth}
  \centering
  \includegraphics[width=\columnwidth]{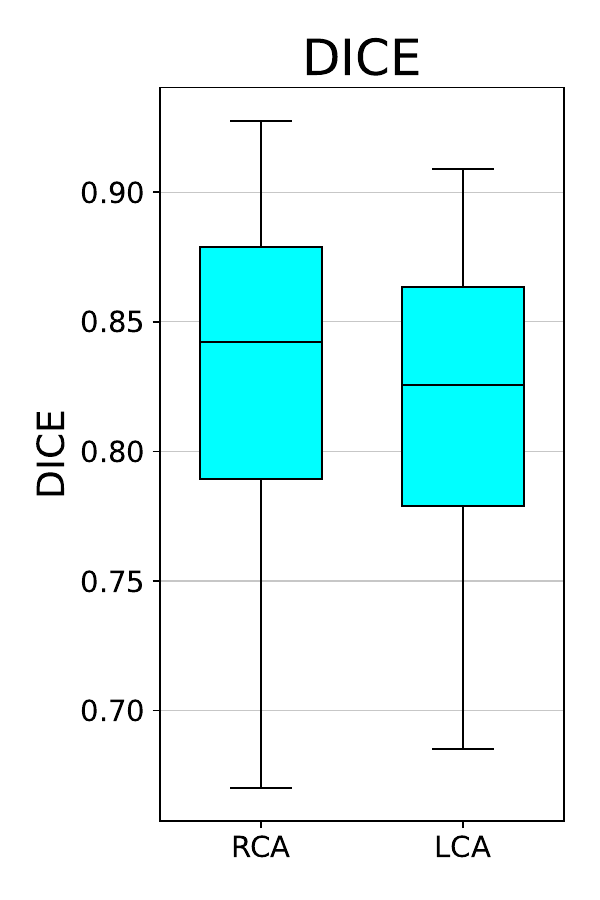}
  \caption{}
\end{subfigure}%
\begin{subfigure}{0.33\columnwidth}
  \centering
  \includegraphics[width=\columnwidth]{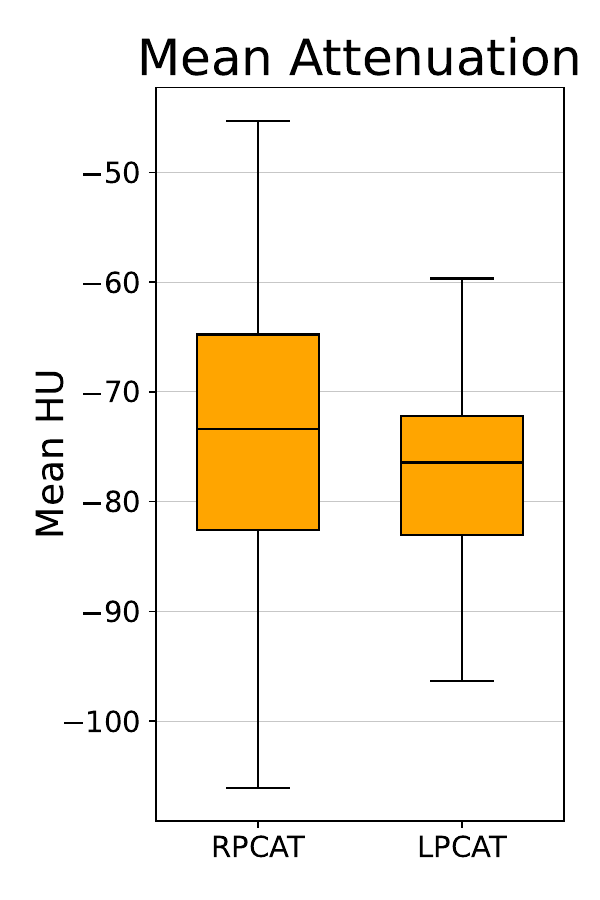}
  \caption{}
\end{subfigure}
\begin{subfigure}{0.33\columnwidth}
  \centering
  \includegraphics[width=\columnwidth]{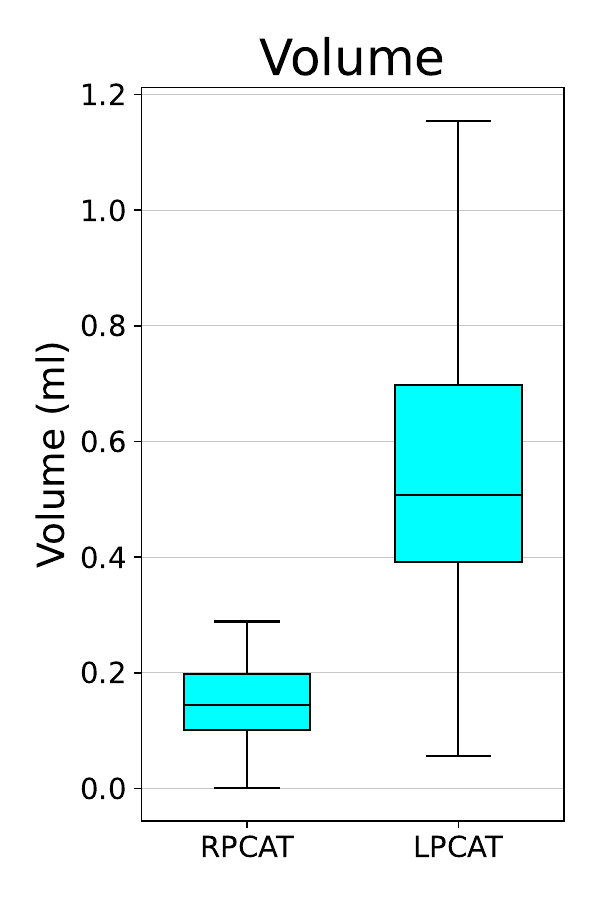}
  \caption{}
\end{subfigure}
\begin{subfigure}{0.49\columnwidth}
  \centering
  \includegraphics[width=\columnwidth]{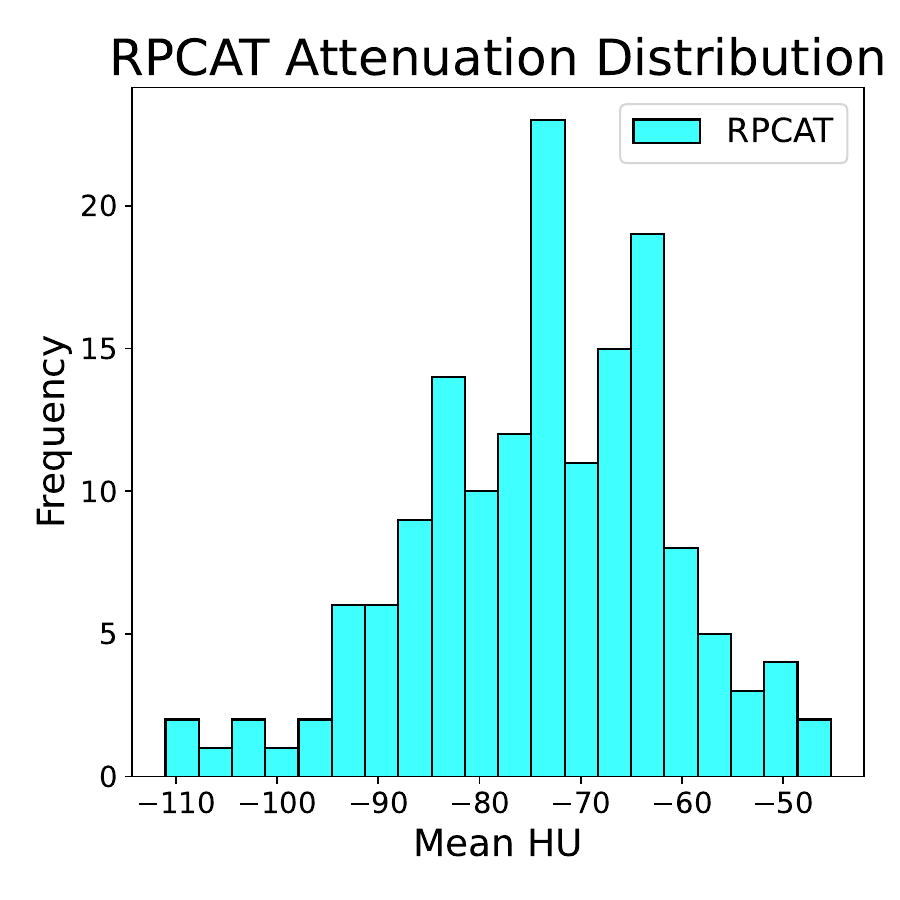}
  \caption{}
\end{subfigure}%
\begin{subfigure}{0.49\columnwidth}
  \centering
  \includegraphics[width=\columnwidth]{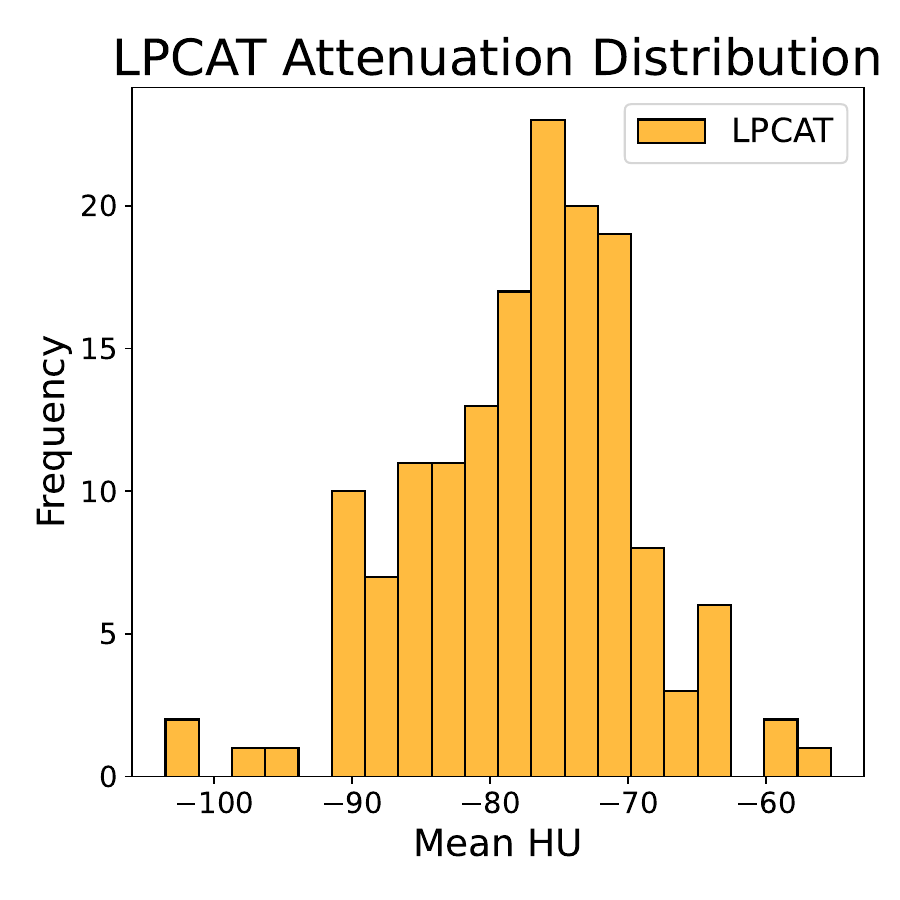}
  \caption{}
\end{subfigure}
\caption{Box plots for: (a) Dice scores for RCA and LCA respectively, (b) Attenuation of the RPCAT and LPCAT regions, and (c) Volume of the RPCAT and LPCAT regions. (d)-(e) HU attenuation distributions for RPCAT and LPCAT.}
\label{fig_results_dice_attenuation_volume_attDistribution}
\end{figure}

The 3D nnUNet model was run on the held-out test split (n = 183 studies). Fig. \ref{fig_money} provides examples of the segmentation output of the model for the left and right arteries. Fig. \ref{fig_results_dice_attenuation_volume_attDistribution}(a) shows box plots of the segmentation dice scores for the left and right coronary arteries respectively. The model achieved median and mean DICE scores of 84\% and 83\% ± 7\% for the RCA, and 83\% and 81\% ± 6\% for the LCA respectively. The quantitative results were similar for both arteries, and are also consistent with prior work \cite{Zeng2023_ImageCAS} on coronary artery segmentation. From a qualitative standpoint, we noticed that the segmentations of the proximal RCA and LCA were successful, but there were some under-segmentations of the smaller distal branches of these arteries. 

\begin{figure}[!h]
\centering
\begin{subfigure}{0.49\columnwidth}
  \centering
  \includegraphics[width=\columnwidth]{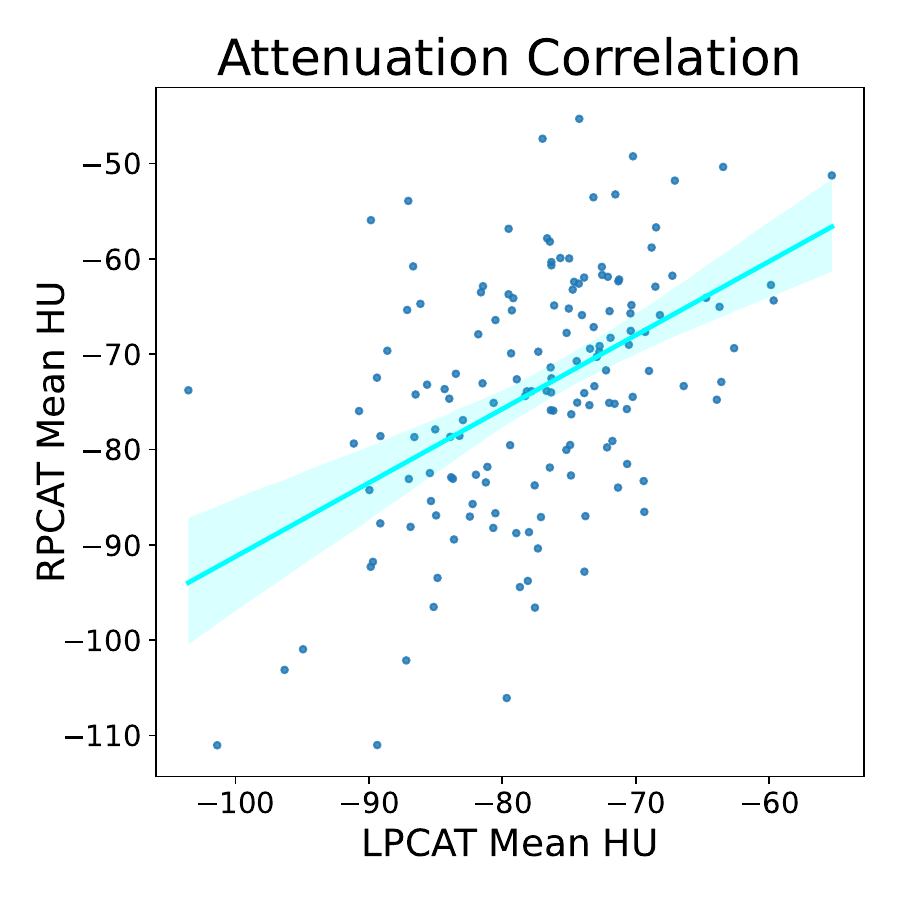}
  \caption{}
\end{subfigure}%
\begin{subfigure}{0.49\columnwidth}
  \centering
  \includegraphics[width=\columnwidth]{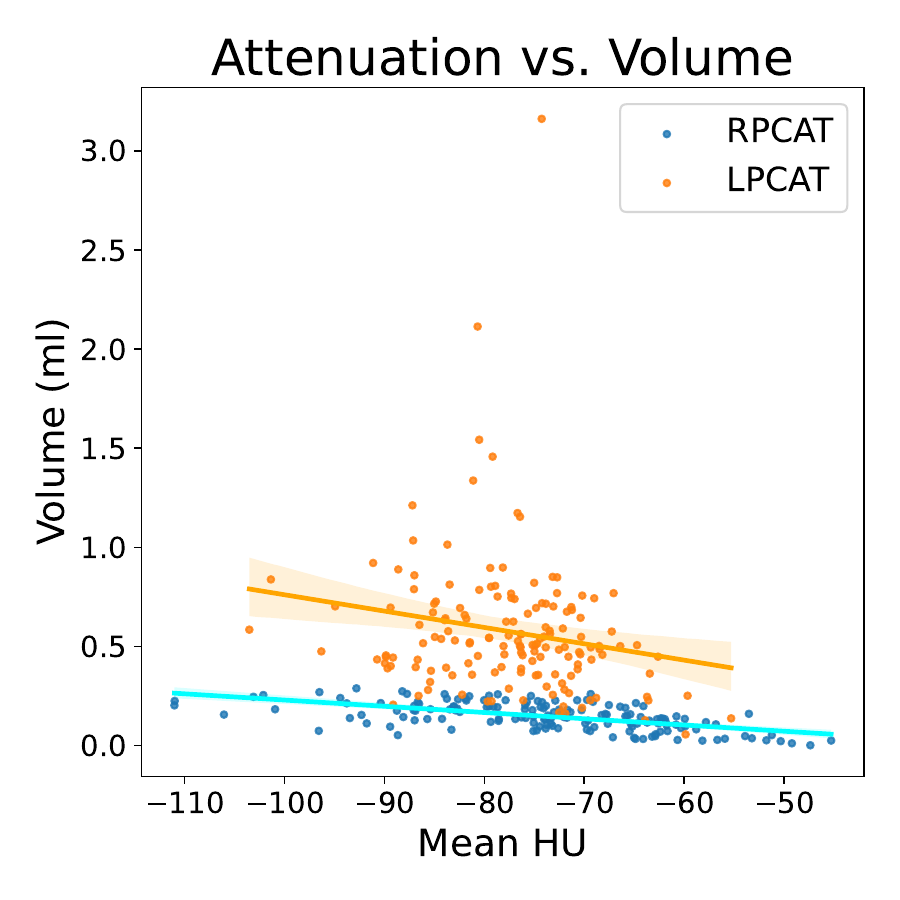}
  \caption{}
\end{subfigure}
\caption{(a) Correlation between the HU attenuations of the LPCAT and RPCAT regions. (b) Correlation between mean attenuation and PCAT volume for RPCAT (cyan) and LPCAT (orange).}
\label{fig_results_line}
\end{figure}

Box plots of the mean attenuation and volume of the LPCAT and RPCAT are also shown in Figs. \ref{fig_results_dice_attenuation_volume_attDistribution}(b) and \ref{fig_results_dice_attenuation_volume_attDistribution}(c). Our results indicated that the RPCAT had a mean attenuation of -73.81 ± 12.69 HU and mean volume of 0.15 ± 0.07 ml, respectively. Similarly, the LPCAT had a mean attenuation of -77.51 ± 7.94 HU and mean volume of 0.58 ± 0.35 ml, respectively. LPCAT mean volume was impacted by the precision of bifurcation point detection. There was a stronger correlation between RPCAT attenuation and volume ($R^2$ = 0.35) than that of LPCAT ($R^2$ = 0.04) as shown in Fig. \ref{fig_results_line}(b). The HU distribution for the RPCAT and LPCAT are also seen in Figs. \ref{fig_results_dice_attenuation_volume_attDistribution}(d) and \ref{fig_results_dice_attenuation_volume_attDistribution}(e). Fig. \ref{fig_results_line}(a) shows the correlation between the mean attenuations of the RPCAT and LPCAT regions with a corresponding $R^2$ = 0.23, indicating a weak relationship between the two measurements.



\section{Discussion}


The 3D nnUNet model segmented the RCA and LCA with mean dice scores of 83\% and 81\% respectively, and the results were consistent with previous work \cite{Zeng2023_ImageCAS}. Our framework was also able to automatically measure the attenuation and volume of the PCAT regions surrounding these arteries. Despite successful segmentation of the coronary arteries, we qualitatively observed under-segmentations of smaller distal arterial branches. However, since we were primarily interested in only the proximal 40 mm segment of the RCA, LAD, and LCX, these under-segmentations had very little to no impact on the PCAT measurements.

It was crucial to find the bifurcation point of the LM as it determined the segmental length for PCAT measurement. Without \textit{a priori} knowledge of the correct segmental length, measurement of PCAT would be difficult and inconsistent. However, differences in anatomy among patients can lead to variations in the length of the LM. This can also have an effect on the detection of the bifurcation, leading to wider standard deviations in the LPCAT volume as seen in Fig. \ref{fig_results_dice_attenuation_volume_attDistribution}(c). 

These challenges were unexplored previously \cite{Chatterjee2022,Xu2020_PCATradiodensity} as a fully automated technique for PCAT measurement had not been used. The automated method in our pilot work can delineate this critical juncture of the LCA and simultaneously use the RCA for PCAT measurement. From Fig. \ref{fig_results_dice_attenuation_volume_attDistribution}(b), the mean attenuation of the LPCAT was fairly close to that of RPCAT. Both hold promise to be used as biomarkers. Importantly, our measurement is repeatable and not subject to annotator biases as in previous semi-automated approaches \cite{Chatterjee2022,Xu2020_PCATradiodensity}.

The mean attenuation for RPCAT that we computed in this work is also consistent with prior work \cite{Ma2023_PCAT,Oikonomou2018_coronaryInflamm,Sugiyama2020,Hoshino2020,Xu2020_PCATradiodensity,Chatterjee2022,Goeller2019,Goeller2021}. The PCAT attenuation at the RCA was higher than that of the LCA by 3.7 HU (-77.5 vs. -73.8). Weak attenuation correlation between RPCAT and LPCAT suggests that the two may have independent inflammatory pathways with regard to coronary artery disease. Although the patients in the ImageCAS dataset had some form of cardiac-related disease, the dataset did not publicly provide the information related to underlying cardiovascular risk and disease status. Therefore, it was not possible to correlate our PCAT measurements against the patient characteristics.

The volume of the PCAT measured in this work may also prove to be beneficial in stratifying patients with cardiac disease, but this measure has not been standardized in the past \cite{Ma2023_PCAT}. We have presented the first method to also standardize the assessment of volumetric PCAT measures. Past literature suggests that the density (attenuation) of fat deposition in the vicinity surrounding the coronary arteries holds more promise as a biomarker \cite{Ma2023_PCAT}. In particular, associating the PCAT mean attenuation with the total burden of plaque and non-calcified plaque in the heart may be of increased diagnostic interest \cite{Ma2023_PCAT,Oikonomou2018_coronaryInflamm,Goeller2019,Goeller2021}. 

To our knowledge, we are the first to develop a fully automated framework for the repeatable measurement of PCAT mean attenuation and volume in the region around the right and left coronary arteries. The method is in contrast to previous semi-automated methods, which are subject to annotator biases. Our framework holds promise for the PCAT measurements to be used as predictive biomarkers for identifying inflammation and the risk of coronary artery disease.

\clearpage

\section{ACKNOWLEDGEMENTS}      

\noindent
This work was supported by the Intramural Research Program of the NIH Clinical Center (project number 1Z01 CL040004). This work utilized the computational resources of the NIH HPC Biowulf cluster.

\section{Compliance with Ethical Standards}

This study was approved by the Institutional Review Board (IRB) at the NIH and performed with retrospectively acquired patient data. The need for informed consent was waived.

\bibliographystyle{IEEEbib}
\bibliography{refs}

\end{document}